\documentstyle[prb,aps]{revtex}

\begin{document}
\draft

\title{Possibility of long-range order in clean mesoscopic cylinders}
\author{M. Lisowski, E. Zipper, and M. Stebelski}
\address{Institute of Physics, University of Silesia,
ul. Uniwersytecka 4, 40-007 Katowice, Poland}
\maketitle

\begin{abstract}
A microscopic Hamiltonian of the magnetostatic interaction is discussed.
This long-range interaction can play an important role in mesoscopic systems
leading to an ordered ground state.

The self-consistent mean field approximation of the magnetostatic
interaction is performed to give an effective Hamiltonian from which the
spontaneous, self-sustaining currents can be obtained.

To go beyond the mean field approximation the mean square fluctuation of the
total momentum is calculated and its influence on self-sustaining currents
in mesoscopic cylinders with quasi-1D and quasi-2D conduction is considered.
Then, by the use of the microscopic Hamiltonian of the magnetostatic
interaction for a set of stacked rings, the problem of long-range order is
discussed. The temperature $T^{*}$ below which the system is in an ordered
state is determined.
\end{abstract}


\section{Introduction}

One of the most exciting areas of physics is the study of mesoscopic
electronic systems, i.e. metal or semiconductor samples which are
sufficiently small and at sufficiently low temperature, such that inelastic
electron-phonon scattering is reduced and the electron propagates as a phase
coherent wave throughout the entire sample \cite{Wash}.

Recently in a series of papers \cite{Wohll,Lis} we discussed a possibility
of spontaneous persistent currents in relatively clean (ballistic regime)
metallic or semiconducting systems of cylindrical geometry.

It was shown in the mean field approximation (MFA) that the inclusion of the
magnetostatic interaction among electrons can lead to an ordered ground
state with spontaneous self-sustaining orbital currents which run without
support of external magnetic field.

In our investigations we considered a collection of many mesoscopic rings
with a thickness $d\ll R$, ($R$ is the radius of the ring) stacked along $z$
axis and the three-dimensional (3D) mesoscopic cylinder of very small
thickness \cite{StebSzopZip}.

In this paper we want to give some justification to the calculation
mentioned above, because there was no microscopic theory of this phenomenon
till now. We will examine a microscopic Hamiltonian for the magnetostatic
(current-current) interaction and show that the self-consistent MFA of it
gives the effective Hamiltonian $H^{MF}$ leading to self-sustaining
currents. We also show that magnetostatic interaction is long-ranged and
therefore the criteria of the MFA are met.

The orbital magnetic interaction and its static version, the magnetostatic
coupling, has been discussed previously by Pines and Nozieres \cite{Pines}
and has been shown to be small in macroscopic metallic samples. However this
interaction should be reexamined in mesoscopic systems where the presence of
energy gaps in the energy spectrum changes qualitatively its physical
properties leading e.g. to persistent currents driven by the static magnetic
flux $\phi $ at low temperatures \cite{Butt}. Persistent currents create
orbital magnetic moments and their interaction can lead to interesting
coherent collective phenomena which bear some resemblance to ferromagnetism
and to superconductivity \cite{Wohll,StebLisZip}.

To go beyond the MFA and discuss fluctuations we follow the ideas developed
by Bloch \cite{Bloch}. He discussed the problem of quantum coherence in a
macroscopic, metallic ring manifested e.g. by the thermodynamically stable
flux trapping at zero external magnetic field. He covers in his paper
different long-range characteristics of the normal and the superconductive
state of a metal.

Irrespective of the specific dynamical properties of the systems he shows
that the general criteria for flux trapping are closely related to the mean
square fluctuation of the total momentum and depend strongly on the
dimensionality of the system.

We will show, using Bloch's formulaes that if we reduce the dimensions of a
macroscopic cylinder made of a normal metal or semiconductor to mesoscopic
dimensions, the system exhibits coherent properties absent in macroscopic
samples. We will formulate the criteria under which flux trapping can be
obtained in mesoscopic cylinders with quasi-1D and quasi-2D conduction. In
particular we will discuss the mean square fluctuation of the total momentum
- the decisive quantity for the characterization of the properties of the
system. We show that it is smaller in mesoscopic systems than in the
corresponding macroscopic ones thus favouring quantum coherence.

Finally, using the microscopic Hamiltonian we discuss the possibility of the
long-range order in a mesoscopic cylinder made of a set of mesoscopic rings.
We calculate the correlation length and the characteristic temperature under
which the system is in a magnetically ordered state.

\section{Magnetostatic interaction}

The general formula for the magnetostatic interaction is of the form:

\begin{equation}
\label{m9}H_{mgt}=-\frac{\mu _0}{8\pi }\int d^3\underline{r}d^3\underline{r}%
^{\prime }\frac{\underline{J}(\underline{r})\underline{J}(\underline{r}%
^{\prime })}{\left| \underline{r}-\underline{r}^{\prime }\right| }, 
\end{equation}
where \underline{$J$}$(\underline{r})=e\underline{p}(\underline{r})/m_e$, 
\underline{$J$}$(\underline{r})$ is the current density, $\underline{p}(%
\underline{r})$ is the momentum of an electron.

Let us assume that the currents run in a set of $M_z$ mesoscopic rings of
small thickness deposited along $z$ axis. We can write

\begin{equation}
\label{m10}\underline{J}(\underline{r})=\sum_{m/1}^{M_z}I_m\oint_{C_m}\delta
^3\left( \underline{r}-\underline{\xi }_m(s)\right) d\underline{\xi }_m, 
\end{equation}
where $C_m$ is given by a parametric equation for an electron going around
the circumference of a ring, $\underline{\xi }_m=\underline{\xi }_m(s)$, $s$
is the coordinate along the circumference of the ring, $I_m$ is the current
in the $m$-th ring.

We obtain

\begin{equation}
\label{m11}H_{mgt}=-\frac 12\sum_{m/1}^{M_z}\sum_{m^{\prime }/1}^{M_z}{\cal L%
}_{mm^{\prime }}I_mI_{m^{\prime }}, 
\end{equation}
where

\begin{equation}
\label{m12}{\cal L}_{mm^{\prime }}=\frac{\mu _0}{4\pi }\oint_{C_m}%
\oint_{C_{m^{\prime }}}\ \frac{d\underline{\xi }_md\underline{\xi }%
_{m^{\prime }}}{\left| \underline{\xi }_m-\underline{\xi }_{m^{\prime
}}\right| },\qquad {\cal L}_{mm^{\prime }}={\cal L}_{m^{\prime }m}. 
\end{equation}

Thus we have obtained the interaction Hamiltonian of the currents from
different rings. In this derivation we neglected the selfinductance effects
in each single ring. It can easily be seen that they are small.

The interaction constant ${\cal L}_{mm^{\prime }}$ depends on the sample
geometry; here it has to be calculated for the rings deposited along $z$
axis at distance $z_{mm^{\prime }}=z_m-z_{m^{\prime }}$.

The result is \cite{Suff}:

\begin{equation}
\label{m13}{\cal L}_{mm^{\prime }}=\mu _0R\left[ \left( \frac
2{h_{mm^{\prime }}}-h_{mm^{\prime }}\right) K-\frac 2{h_{mm^{\prime
}}}E\right] , 
\end{equation}
where

\begin{equation}
h_{mm^{\prime }}^2=\frac{4R^2}{4R^2+z_{mm^{\prime }}^2}, 
\end{equation}

\begin{equation}
\label{m14}K=\int_0^{\pi /2}\frac{d\theta }{\left( 1-h^2\sin ^2\theta
\right) ^{1/2}},\ E=\int_0^{\pi /2}\left( 1-h^2\sin ^2\theta \right) d\theta
,
\end{equation}
$K$ and $E$ are the elliptical integrals of the $I$ and $II$ kind
respectively.

\ 

The $z$ dependence of the coupling constant ${\cal L}$ is presented
in Fig. 1. We see that the interaction (\ref{m11}) is a long-range
interaction. For small $z$ it falls down slowly proportionally to $\mu
_0R\ln \left( R/z-2\right) $, for large $z$ it falls down faster
proportionally to $1/z^3$. The interaction constant depends only on $R$
and on the relative distance of the centres of the rings.

As the currents $I_m$ can run only in the clockwise or anticlockwise
direction the Hamiltonian (\ref{m11}) has the form of the Ising Hamiltonian.
It can be also expressed via the momenta $p_m$, $p_{m^{\prime }}$ from
different rings.

For the ring geometry we get

\begin{equation}
\label{m16}I_m=\frac e{2\pi Rm_e}p_m, 
\end{equation}
$$
p_m=\sum_{n/1}^{N^R}p_{nm}, 
$$
where $N^R$ is a number of conducting electrons in a single ring.

$H_{mgt}$ given by Eq. (\ref{m11}) can be rewritten in the form:

\begin{equation}
\label{m17}H_{mgt}=-\frac{e^2}{2m_e^2}\sum_{m/1}^{M_z}\sum_{m^{\prime
}/1}^{M_z}g_{mm^{\prime }}p_mp_{m^{\prime }}, 
\end{equation}

$$
g_{mm^{\prime }}=\frac 1{4\pi ^2R^2}{\cal L}_{mm^{\prime }}. 
$$

If we add to $H_{mgt}$ the kinetic energy term and if we assume that the
external magnetic field parallel to the $z$ axis is applied to the system,
we obtain Hamiltonian $H$: 
\begin{equation}
\label{m19}H=\frac 1{2m_e}\sum_{m/1}^{M_z}\sum_{n/1}^{N^R}p_{nm}^2-\frac{e^2%
}{2m_e^2}\sum_{m/1}^{M_z}\sum_{m^{\prime }/1}^{M_z}g_{mm^{\prime
}}p_mp_{m^{\prime }}, 
\end{equation}
where $p_{nm}=p_{nm}^0-eA_e$; $p_{nm}^0$ is the momentum of the $n$-th
electron in $m$-th ring. $A_e$ is the vector potential of an external
magnetic field which points in the $x$ direction measured along the ring.

The first term in Eq. (\ref{m19}) represents the kinetic energy of electrons
in the external magnetic field, the second term represents the orbital
magnetic interaction; due to the negative sign it favours states with large
total momentum in competition with the kinetic energy.

\section{Mean field approximation of magnetostatic coupling}

We have seen in chapter 2 that the magnetostatic coupling is a long-range
interaction. This indicates in particular that thermodynamic fluctuations of
the current will be strongly supressed \cite{Fisher}.

Let's perform a self-consistent MFA of the interaction (\ref{m17}), such
approximation is known to be good for a long-range interaction.

\begin{equation}
\label{m20}H_{mgt}=-\frac{e^2}{2m_e^2}\sum_mp_m\sum_{m^{\prime
}}g_{mm^{\prime }}p_{m^{\prime }}\equiv -\frac e{2m_e}\sum_mp_mA_I(z_m), 
\end{equation}

\begin{equation}
\label{m21}A_I(z_m)=\frac e{2m_e}\sum_{m^{\prime }}g_{mm^{\prime
}}p_{m^{\prime }}, 
\end{equation}
$$
BC\rightarrow B\left\langle C\right\rangle +\left\langle B\right\rangle
C-\left\langle C\right\rangle \left\langle B\right\rangle 
$$

\begin{equation}
\label{m22}H_{mgt}=-\frac e{2m_e}\sum_m\left[ 2p_m\left\langle
A_I(z_m)\right\rangle -\left\langle p_m\right\rangle \left\langle
A_I(z_m)\right\rangle \right] , 
\end{equation}
where the first term in Eq. (\ref{m22}) has been obtained by use of the
symmetry relation (\ref{m12}), $\left\langle A_I(z_m)\right\rangle
=(e/2m_e)\sum_{m^{\prime }}g_{mm^{\prime }}\left\langle p_{m^{\prime
}}\right\rangle \equiv A_I$ where we assumed that $\left\langle
p_m\right\rangle \equiv \left\langle p\right\rangle $ is the same for all
rings, and hence from Eq. (\ref{m16}) $\left\langle I_m\right\rangle \equiv
\left\langle I\right\rangle $.

Assuming that our stack of $M_z$ rings forms a long cylinder of length $l$
we can calculate the vector potential $A_{I.}$ We get 
\begin{equation}
\label{m23}A_I=\mu _0R\frac{M_z\left\langle I\right\rangle }{2l}. 
\end{equation}
Calculating the current $\left\langle I\right\rangle $ with a total vector
potential $A=A_e+A_I$ we get

\begin{equation}
\label{m24}\left\langle I\right\rangle =\frac e{2\pi Rm_e}\left(
\left\langle p\right\rangle -N^ReA_I\right) . 
\end{equation}
Inserting (\ref{m24}) into (\ref{m23}) we obtain the self-consistent
equation for $A_I:$

\begin{equation}
A_I=\frac{\mu _0eM_z}{4\pi lm_e}\left( \left\langle p\right\rangle
-N^ReA_I\right) , 
\end{equation}
from which we get

\begin{equation}
\label{m27}eA_I=\frac \eta {1+\eta N^RM_z}M_z\left\langle p\right\rangle , 
\end{equation}
where%
$$
\eta =\frac{\mu _0e^2}{4\pi lm_e}. 
$$
Inserting $\left\langle p\right\rangle $ from Eq. (\ref{m27}) into Eq. (\ref
{m22}) we obtain the Hamiltonian $H$ (Eq.(\ref{m19})) in the self-consistent
mean field approximation:

\begin{equation}
\label{m28}H^{MF}=\frac 1{2m_e}\sum_{m/1}^{M_z}\sum_{n/1}^{N^R}\left(
p_{nm}-eA_I\right) ^2+\frac{\phi _I^2}{2{\cal L}}, 
\end{equation}
where ${\cal L}=\mu_0\pi R^2 (\sqrt{l^2 - R^2}-R)/l^2$, $\phi_I=2\pi RA_I$.

The Hamiltonian (\ref{m28}) was the basis of our previous investigations 
\cite{Wohll} of spontaneous self-sustaining currents (see Eq. (\ref{pII}), (%
\ref{pIII})). Its derivation from the long-range current-current interaction
serves as a justification of the use of the Hamiltonian (\ref{m28}) to
investigate magnetic properties of mesoscopic systems.

It has been generally believed that the MFA should work well for long-range
forces. However, it has been shown in Ref. \cite{Fried} that the above
statement is correct if an additional condition is fulfilled. The authors
defined there the quantity $S$: 
\begin{equation}
\label{mS}S=\frac{\left( \sum_{m^{\prime }}{\cal L}_{mm^{\prime }}\right) ^2%
}{\sum_{m^{\prime }}{\cal L}_{mm^{\prime }}^2},
\end{equation}
where ${\cal L}_{mm^{\prime }}$ is the interaction constant. They proved
that MFA is correct if $S\gg 1$.

We have calculated $S$ with ${\cal L}_{mm^{\prime }}$ given by Eq. (\ref{m13}%
) for the following set of parameters: $b\equiv z_{m,m+1}=10$\AA , $R=5000$%
\AA , $M_z\sim 10^3$. We have obtained $S\sim 10^5$; it means that the MFA
should work well in the case considered by us.

\section{Summary of the Bloch's results}

We are going now to investigate the possibility of flux trapping in
mesoscopic cylinders using the ideas developed by Bloch \cite{Bloch}. The
general criteria for flux trapping are closely related to mean square
fluctuations and give a natural way to describe the system in terms of the
two-fluid model. At first we briefly recall Bloch's results.

Let us consider a system of $N$ particles with mass $m$ and charge $e$,
contained in a ring with radius $R$ and radial width $d\ll R$. We assume
that the magnetic field parallel to $z$ axis is caused by a current around
the ring, so the vector potential $A=\phi /(2\pi R)$ points in the $x$
direction.

The total momentum in the $x$ direction is given by 
\begin{equation}
\label{b1}P_x=\sum_{n/1}^Np_n^0, 
\end{equation}
where $p_n^0$ represents the momentum of $n$-th particle ($n=1,2,...,N$).

The Hamiltonian of the particles is of the form: 
\begin{equation}
\label{b2}H=\frac{\left( P_x-NeA\right) ^2}{2Nm_e}+H^{\prime }, 
\end{equation}
where $H^{\prime }$ contains the kinetic energy of the motion in the $y$ and 
$z$ direction and of the relative motion in the $x$ direction as well as any
additional terms which arise from interactions and characterize the specific
dynamical properties of the system.

From the symmetries and periodic boundary conditions \cite{Bloch} we get the
eigenvalues of $P_x$: 
\begin{equation}
\label{b3}P=n\frac \hbar R=(N\nu +\mu )\frac \hbar R, 
\end{equation}
$n$, $\nu $ are arbitrary integers, $\mu $ is likewise an integer such that%
$$
-\frac N2<\mu \leq \frac N2. 
$$

Eigenenergies of the system are given by 
\begin{equation}
\label{b4}E_{\mu \nu q}=\frac{\hbar ^2N\left( \nu -\phi ^{\prime }+\frac \mu
N\right) ^2}{2m_eR^2}+E_{q\mu }^{\prime }, 
\end{equation}
$q$ represents the system of additional quantum numbers necessary in
addition to $P$ $(\nu ,\mu )$ in order to fully characterize the state of
the system; $\phi ^{\prime }=\phi /\phi _0$, $\phi _0=h/e.$

Using Eq. (\ref{b4}) we can calculate the free energy from the particles.
The flux dependent part of the total free energy is 
\begin{equation}
\label{b12}F(\phi ^{\prime })=F_1(\phi ^{\prime })+F_2(\phi ^{\prime }), 
\end{equation}
where $F_1(\phi ^{\prime })$ is a periodic function of $\phi ^{\prime }$
with period 1, 
\begin{equation}
\label{b10}F_1(\phi ^{\prime })=-k_BT\ln Z_1(\phi ^{\prime }), 
\end{equation}
\begin{equation}
\label{b14}Z_1(\phi ^{\prime })=\sqrt{\frac \pi {N\gamma }}\left(
1+2\sum_{g/1}^\infty a_ge^{-\frac{(\pi g)^2}{N\gamma }}\cos 2\pi g\phi
^{\prime }\right) , 
\end{equation}
\begin{equation}
\label{b15}a_g=\sum_\mu z_\mu e^{-2\pi ig\mu /N}, 
\end{equation}
$\gamma =\hbar ^2/\left( 2m_eR^2k_BT\right) $; $z_\mu $ is the statistical
weight of the state $E_{q\mu }^{\prime }$ , 
\begin{equation}
\label{b7}\sum_\mu z_\mu =1,\ z_\mu \geq 0. 
\end{equation}
\begin{equation}
\label{b11}F_2(\phi ^{\prime })=\frac{\hbar ^2\phi ^{\prime 2}}{2e^2{\cal L}}%
, 
\end{equation}
$F_2(\phi ^{\prime })$ is the energy stored in the magnetic field, ${\cal L}$
is the self-inductance of the ring.

Thermodynamically stable flux trapping is determined by those values of $%
\phi ^{\prime }$ for which $F(\phi ^{\prime })$ has a minimum. To achieve it
a strong variation of $F_1(\phi ^{\prime })$ is necessary to prevent the
dominance of $F_2(\phi ^{\prime })$ which has minimum at $\phi ^{\prime }=0$.

Let us consider now three special cases which elucidate how different states
of matter can be described in this model. The information about specific
properties of the system are contained in the quantities $z_\mu $.

{\bf A. }$z_\mu $ is independent of $\mu $ , i.e. $z_\mu =1/N$.

After some algebra we arrive at 
\begin{equation}
\label{b17}Z_1(\phi ^{\prime })=\sqrt{\frac \pi {N\gamma }}\left(
1+2\sum_{g/1}^\infty e^{-\frac{(\pi g)^2N}\gamma }\cos 2\pi gN\phi ^{\prime
}\right) . 
\end{equation}
The general periodicity in $\phi ^{\prime }=1$ is accompanied here by a far
shorter period $\phi ^{\prime }=1/N$ what is vanishingly small for large $N$%
. Besides the amplitude of the oscillation is small at any realistic
temperature leading to a very small variation of $F_1(\phi ^{\prime })$ and
hence to the absence of stable flux trapping. This situation corresponds to
the case where the system exhibits no long-range order and is characteristic
of the normal state of a metal.

{\bf B. }$z_\mu =\delta _{\mu 0}.$

From Eq. (\ref{b14}) we find: 
\begin{equation}
\label{b18}Z_1(\phi ^{\prime })=\sqrt{\frac \pi {N\gamma }}\left(
1+2\sum_{g/1}^\infty e^{-\frac{(\pi g)^2}{N\gamma }}\cos 2\pi g\phi ^{\prime
}\right) , 
\end{equation}
which has the periodicity with $\phi ^{\prime }=1$. However in order to get
a stable flux trapping for $\phi ^{\prime }\neq 0$ we need 
\begin{equation}
\label{b19}N\gamma =\frac{N\hbar ^2}{2m_eR^2k_BT}\gg 1, 
\end{equation}
what is satisfied at low temperatures $T$. Then $F_1(\phi ^{\prime })$ can
well dominate the part $F_2(\phi ^{\prime })$ in Eq. (\ref{b12}). The
pronounced minima of $F(\phi ^{\prime })$ occur at $\phi ^{\prime }=\nu $
and are equivalent with stable flux trapping. This case corresponds to a
condensed ideal Bose gas where all particles have momentum $p=\nu \hbar /R$.

{\bf C. }$z_\mu =\frac 12\left( \delta _{\mu 0}+\delta _{\mu \frac
N2}\right) .$

The partition function is of the form: 
\begin{equation}
\label{b21}Z_1(\phi ^{\prime })=\sqrt{\frac \pi {N\gamma }}\left(
1+2\sum_{g/1}^\infty e^{-\frac{(2\pi g)^2}{N\gamma }}\cos 4\pi g\phi
^{\prime }\right) . 
\end{equation}
This expression has the periodicity in $\phi ^{\prime }$ with $\phi ^{\prime
}=\frac 12$ i.e. with $\phi =h/2e$. This case exhibits the property that $%
N/2 $ pairs have all the same momentum $p=\nu \hbar /R$ and corresponds to
the long-range order characteristic of a superconductor at temperatures in
which the condition (\ref{b19}) is satisfied.

\ 

To consider other, more general cases one assumes that the total momentum $P$
can admit the values given by Eq. (\ref{b3}) with $\mu \neq 0$, but with a
sharp maximum around $\mu =0$. The formula for $Z_1(\phi ^{\prime })$ has
then the form: 
\begin{equation}
\label{b25}Z_1(\phi ^{\prime })=\sqrt{\frac \pi {N\gamma }}\left(
1+2\sum_{g/1}^\infty e^{-\frac{(\pi g)^2}{N\gamma (1-\rho )}}\cos 2\pi g\phi
^{\prime }\right) , 
\end{equation}
where 
\begin{equation}
\label{b26}\rho =\frac{\left\langle \left( \Delta P\right) ^2\right\rangle }{%
Nm_ek_BT},\qquad 0\leq \rho \leq 1, 
\end{equation}
$\left\langle \left( \Delta P\right) ^2\right\rangle $ is the mean square
fluctuation of momentum $P$ around the set of values $\phi ^{\prime }N\hbar
/R$, where $\phi ^{\prime }$ is an integer multiple of the period, $\rho $
can be called a relative fluctuation. 
\begin{equation}
\label{b24}\left\langle \left( \Delta P\right) ^2\right\rangle =\frac{\hbar
^2}{R^2}\left\langle \left( \Delta \mu \right) ^2\right\rangle
=-m_ek_BT\sum_\alpha p_\alpha \frac{\partial f(p_\alpha )}{\partial p_\alpha 
}, 
\end{equation}
where $\left\langle \left( \Delta P\right) ^2\right\rangle =\left\langle
P^2\right\rangle -\left\langle P\right\rangle ^2$ ; $f(p_\alpha )$ is the
mean number of particles with momentum $p_\alpha =\hbar \left( \alpha -\phi
^{\prime }\right) /R$, $\alpha =0,\pm 1,\pm 2,...$

Eq. (\ref{b25}) covers the special cases discussed before. The case {\bf A}
corresponds to the situation where all values of $\mu $ are equally probable
what leads to $\rho \rightarrow 1$; $Z_1(\phi ^{\prime })$ becomes then
independent of $\phi ^{\prime }$ with the exclusion of flux trapping. Cases 
{\bf B} and {\bf C} correspond to $\rho =0$ and thus describe the systems in
the coherent state with no fluctuations.

\ 

The intermediate case: 
\begin{equation}
\label{b27}0<\left\langle \left( \Delta P\right) ^2\right\rangle <Nm_ek_BT 
\end{equation}
or equivalently $0<\rho <1$ still results in a pronounced variation of $%
F_1(\phi ^{\prime })$ as long as 
\begin{equation}
\label{b28}N\gamma (1-\rho )\gg 1. 
\end{equation}
It means that stable flux trapping can be expected as soon as $\left\langle
\left( \Delta P\right) ^2\right\rangle $ is found to be smaller than the
maximal value of mean square fluctuation of $P$.

\section{Flux trapped in mesoscopic cylinders}

We will perform now, using Bloch's formalism, some model calculations of
flux trapped in mesoscopic hollow cylinders of radius $R$, length $l$ and
wall thickness $d$ ($d\ll R$) made of a normal metal or semiconductor. Such
cylinders can be treated as a multichannel system with $M_z$ channels in the
length and $M_r$ channels in the thickness of the cylinder ($M\equiv M_zM_r$%
), with the total number of conducting electrons $N=N^RM$. We will study the
systems with quasi-1D and quasi-2D conduction. It is known that coherent
response in mesoscopic cylinders can be obtained \cite{IBM,StebSzopZip}
for systems with large phase correlation of currents from
different channels which is related to the shape of the Fermi Surface (FS).
The most favourable situation is for systems with quasi-1D conduction, i.e.
with a flat FS parallel to axes of the wave vector \underline{$k$}. There
exists then a perfect corelation among the channel currents and the magnetic
response is the strongest. Such FS can be obtained in bcc crystals \cite
{Dzialo}, in low dimensional organic conductors with the overlap of the
orbitals mainly in one direction, and when a cylinder is made of a set of $M$
quasi-1D rings stacked along $z$ axis by e.g. lithographic method. Such
multiple quantum chains can be mapped into a system with the flat
(rectangular) FS.

In the case of quasi-2D conduction our cylinder can be constructed from a
set of $M_r$ 2D coaxial cylinders (e.g. multiwall carbon nanotubes and
cylinders made of a material with layered structure).

To simulate different shapes of the 2D FS \cite{StebSzopZip,Crack}
we can use the equation: 
\begin{equation}
\label{kFn}k_F^u=k_{F_x}^u+k_{F_z}^u, 
\end{equation}
where $u$ is an integer number.

For $u=2$ we get the circular FS, for $u\geq 12$ the rectangular one and for 
$2<u<12$ the rectangular FS with rounded corners. For the circular FS the
currents from different channels add almost without correlation, the
correlation increases with increasing the curvature of the FS, i.e. with
increasing $u$.

Let us consider at first a set of $M$ quasi-1D rings stacked along $z$ axis
(or in general a cylinder with quasi-1D conduction). Let us assume that the
magnetic field is caused by persistent currents from all rings. Thus we meet
conditions from the Bloch's paper and our mean field Hamiltonian $H^{MF}$
(Eq. (\ref{m28})) leads to the free energy given by Eq. (\ref{b12}) with 
\begin{equation}
F_1(\phi ^{\prime })=MF_1^R(\phi ^{\prime }), 
\end{equation}
where $F_1^R(\phi ^{\prime })$ is the free energy of a single ring.

In general an external magnetic flux $\phi _e$ parallel to the $z$ axis can
be also applied to the system but we are mainly interested in the
self-sustaining flux at $\phi _e=0$.

The necessary condition for coherent behaviour has the form: 
\begin{equation}
N^R\gamma \equiv \frac{\Delta _0}{k_BT}\gg 1, 
\end{equation}
where $\Delta _0=\hbar ^2N^R/(2m_eR^2)$, $\Delta _0$ is the quantum size
energy gap at the FS.

Let us calculate the momentum $P$ defined as in Eq. (\ref{b3}) for three
different model cases.

{\bf 1.} If the number of electrons $N^R$ in each ring is odd we find:\label
{11} 
\begin{equation}
P=N\nu \frac \hbar R\quad \text{for }\left( \nu -\frac 12\right) <\phi
^{\prime }<\left( \nu +\frac 12\right),\quad \nu =0,1,... 
\end{equation}
what corresponds to the case {\bf B} with $z_\mu =\delta _{\mu 0}$ and $\rho
=0$.

Calculating the minimum of the total free energy (Eq. (\ref{b12})) at $\phi
_e=0$ and at $T\ll \Delta _0/k_B$ we get the value of the flux trapped $\phi
^{\prime t}$ in the cylinder \cite{Wohll}: 
\begin{equation}
\label{p4}\phi _{odd}^{\prime t}=\frac \nu {1+\kappa }, 
\end{equation}
where%
$$
\kappa =\frac{4\pi ^2m_eR^2}{e^2{\cal L}N}\quad \text{and}\quad |\nu |\leq
\frac 12\left( 1+\kappa ^{-1}\right) . 
$$
Eq. (\ref{p4}) reveals the influence of the finite size effect - the flux
trapped is quantized in units less than $\phi _0$ \cite{Bardeen}. For
macroscopic samples $\kappa \rightarrow 0$ and $\phi _{odd}^{\prime t}=\nu .$

{\bf 2.} If the number of electrons $N^R$ in each ring is even, the momentum 
$P$ is 
\begin{equation}
P=N\left( \nu +\frac 12\right) \frac \hbar R\quad \text{for }\left( \nu
-\frac 12\right) <\phi ^{\prime }<\left( \nu +\frac 12\right),\quad \nu
=0,1,... 
\end{equation}
It corresponds to $z_\mu =\delta _{\mu \frac N2}$ and stable values of
trapped flux are 
\begin{equation}
\phi _{even}^{\prime t}=\frac{\nu +\frac 12}{1+\kappa },\quad \left| \nu
+\frac 12\right| <\frac 12\left( 1+\kappa ^{-1}\right) . 
\end{equation}
For macroscopic samples $\phi _{even}^{\prime t}=\nu +\frac 12$ what
corresponds to stable minima of the free energy at half-integral values of $%
\phi _0$.

We see that a system under consideration exhibits thermodynamically stable
persistent currents and flux trapped at temperatures $T\ll \Delta _0/k_B$.
This condition is easily satisfied for mesoscopic rings at $T\leq 1K$ (e.g.
for $R\sim 1$ $\mu $m$,$ $\Delta _0/k_B\sim 12.5K$), it is however
unrealistic for macroscopic samples (for $R\sim $1 cm, $\Delta _0/k_B\sim
10^{-3}K$) \cite{Schick}.

Our treatment is not only for identical rings. We have performed the
calculations for a set of rings in which the number of electrons $N^R$
changes in the range $N^R=\bar{N}^R \pm \Delta N^R$, $\Delta N^R=10$.
We have considered two kinds of changes:

$i$) $N^R$ fluctuates from $\bar{N}^R$ to $\bar{N}^R \pm 2n$, $n=1,2,3,4,5$
- the influence of such fluctuation on the persistent current is very small.

$ii$) $N^R$ changes from $\bar{N}^R$ to $\bar{N}^R \pm (2n-1)$
- the influence of this kind of changes on the current is pretty large
(see below).

In the presented paper we study mainly the systems with the diamagnetic
reaction on small magnetic flux $\phi$. The rings with odd $N^R$ give
a diamagnetic current whereas those with even $N^R$ give a paramagnetic
current. We have found that the diamagnetic reaction and trapped flux
can still be obtained if about 20\% of rings carry an even number of
conducting electrons.

{\bf 3.} Finally, we may also consider the model case where roughly a half
of the rings has an even number of electrons and a half of the rings has an
odd number of electrons. It corresponds to the case {\bf C} where $z_\mu
=\frac 12\left( \delta _{\mu 0}+\delta _{\mu \frac N2}\right) $ and the
minima of the free energy occur with the twice smaller period.

We have not discussed this situation here in details because the total
current in this case is paramagnetic \cite{Wohll}.

\ 

We are going now beyond the MFA and consider fluctuations around the set of $%
P$ values discussed above. We assume that the total momentum $P$ in a
cylinder given by Eq. (\ref{b3}) can admit the values with $\mu \neq 0$
and/or $\mu \neq N/2$ but with a maximum at $\mu =0$ and/or $\mu =N/2$. This
assumption leads to the partition function $Z_1(\phi ^{\prime })$ given by
Eq. (\ref{b25}) and the criteria for quantum coherence manifesting
themselves in the flux trapping are ultimately related to the magnitude of
the mean square fluctuation of momentum $P$. The condition for coherent
behaviour takes then the form: 
\begin{equation}
\frac{\Delta _0(1-\rho )}{k_BT}\gg 1, 
\end{equation}
and we see that the fluctuations decrease the energy gap. Eq. (\ref{b25})
covers also as a special case a situation characteristic of a normal state
of a metal (no flux trapping) in which all values of $\mu $ can be found,
with equal probability, in the formula for $P$.

Stable flux trapping can be expected as soon as $\left\langle \left( \Delta
P\right) ^2\right\rangle $ is found to be a small fraction below $Nm_ek_BT$,
called by Bloch the equipartition value. In the following we will calculate $%
\left\langle \left( \Delta P\right) ^2\right\rangle $ from Eq. (\ref{b24})
and relate its magnitude to the coherent behaviour of the sample.

The mean number of electrons with momentum $p_{\alpha m}$, $f(p_{\alpha m})$
is given by the Fermi-Dirac distribution function: 
\begin{equation}
f(p_{\alpha m})\equiv f(E_{\alpha m}(\phi ))=\frac 1{e^{(E_{\alpha m}-\mu
_c)/k_BT}+1}, 
\end{equation}
where $\mu _c$ is calculated from the condition: 
\begin{equation}
N=M_r\sum_{m/1}^{M_z}\sum_{\alpha /0,\pm 1}^{\pm \infty }f(E_{\alpha m}(\phi
)). 
\end{equation}
The electron energy eigenvalues, calculated by the use of periodic boundary
conditions in the $x$ direction and cyclic boundary conditions in the $z$
direction: 
\begin{equation}
\label{pI}E_{\alpha m}=\frac 1{2m_e}\left[ \left( \frac \hbar R\alpha
-eA\right) ^2+\hbar ^2k_z^2(m)\right] ,\qquad A=\frac \phi {2\pi R}, 
\end{equation}
\begin{equation}
\label{pII}\phi =\phi _e+\phi _I, 
\end{equation}
$\phi $ is the total flux contained in the cylinder, $\phi _I={\cal L}I(\phi
)$, where

\begin{equation}
\label{pIII}I(\phi )=\frac{e\hbar }{2\pi m_eR^2}M_r\sum_{m/1}^{M_z}\sum_{%
\alpha /0,\pm 1}^{\pm \infty }(\alpha -\phi ^{\prime })f(E_{\alpha m}(\phi
)),
\end{equation}
and after expansion in the Fourier series \cite{StebSzopZip}:%
$$
I(\phi )=\frac{4eak_BT}{\pi \hbar }M_r\sum_{m/1}^{M_z}\sum_{g/1}^\infty
k_{F_x}(m)\frac{\exp \left( -\frac{2\pi ^2gk_BT}{\Delta _0(1-\rho )}\right) 
}{1-\exp \left( -\frac{4\pi ^2gk_BT}{\Delta _0(1-\rho )}\right) }\cos (2\pi
gRk_{F_x}(m))\sin (2\pi g\phi ^{\prime }), 
$$
where $a$ is the lattice constant, and according to Eq. (\ref{kFn}) $%
k_{F_x}(m)=k_F\left[ 1-\left( k_z(m)/k_F\right) ^u\right] ^{1/u}$, $%
k_z(m)=m\pi /l$, $m=1,2,...,M_z$.

Eqs. (\ref{pII}) and (\ref{pIII}) form a set of self-consistent equations
for the current. The question of existence of self-sustaining, persistent
currents is reduced to the problem whether these equations have stable,
nonvanishing solutions at $\phi _e=0$.

Notice that the dispertion relation (\ref{pI}) is modified by the presence
of the flux $\phi _I$ coming from the currents and has to be calculated in a
self-consistent way. We show below that $\phi _I$ produces a dynamic gap in
the system and therefore increases coherence.

The values of $\left\langle \left( \Delta P\right) ^2\right\rangle $ for
different shapes of the FS and for $R\simeq 10^4$\AA $/(2\pi )$, $M_z=10000$%
, $M_r=100$ at $T=15K$ are presented in Table 1.

For the rectangular FS, corresponding to quasi-1D conduction, we assumed
that it lies in the middle of an energy gap $\Delta _0$ for an electron
going along the circumference of the cylinder. We see that the magnitude of $%
\left\langle \left( \Delta P\right) ^2\right\rangle $ (or the corresponding
relative fluctuation $\rho $) decreases with increasing the curvature of the
FS. The difference between the normal and the coherent state of a mesoscopic
cylinder is reflected in the magnitude of $\left\langle \left( \Delta
P\right) ^2\right\rangle .$

In the ideal, limiting case corresponding to $\left\langle \left( \Delta
P\right) ^2\right\rangle =0$ (case {\bf 1} on page \pageref{11}) the system
is fully coherent i.e. $N=N_c$, where $N_c$ is the number of electrons in a
coherent state.

Finite values of $\left\langle \left( \Delta P\right) ^2\right\rangle $ can
be interpretated by use of the two-fluid model \cite{Bloch,StebLisZip}
as being proportional to $N_n=N-N_c$, $N_n$ is the number of
particles in the normal state. We can write $\left\langle \left( \Delta
P\right) ^2\right\rangle =N_nm_ek_BT$.

The maximal values of $\left\langle \left( \Delta P\right) ^2\right\rangle $
given by Eq. (\ref{b24}) will be obtained for macroscopic values of $R$
where the replacement of the sums over $\alpha $ by integrals is permitted.
This gives us $\left\langle \left( \Delta P\right) ^2\right\rangle =Nm_ek_BT$
or $N=N_n$ and the system is in a normal phase. However for mesoscopic
cylinders such replacement is not allowed and the presence of finite size
energy gaps leads to $N_n<N$ what means that a part of the electrons is in a
coherent state. The presence of coherent electrons results in persistent
currents, the amplitude of which depends on the shape of the FS.

Persistent currents driven by an external flux $\phi _e$ vanish if we switch
the external field off. However the presence of the flux $\phi _I$ coming
from the magnetostatic interaction can lead to persistent self-sustaining
currents or in other words to flux trapped. In general persistent currents
can be paramagnetic or diamagnetic. Paramagnetic self-sustaining currents
correspond to spontaneous currents \cite{StebSzopZip}, and diamagnetic
self-sustaining currents correspond to flux trapping \cite{StebLisZip}. In
the following we will study mainly the diamagnetic solutions.

In order to discuss the influence of $\phi _I$ on the coherent properties of
mesoscopic cylinders we calculate an energy gap at the FS for electrons
going around the circumference of the cylinder (and for $\phi <\phi _0/2$).
Using Eq. (\ref{pI}) we find: 
\begin{equation}
\Delta _F\equiv E_{\alpha _F+1,m}-E_{\alpha _F,m}=\Delta _0\left( 1-2\phi
_e^{\prime }+2\frac{{\cal L}\left| I\right| }{\phi _0}\right) . 
\end{equation}
$\Delta _F$ contains a term 
\begin{equation}
\Delta _d\equiv \Delta _0\frac{{\cal L}\left| I\right| }{\phi _0}, 
\end{equation}
$\Delta _d$ is the dynamic part of an energy gap which should increase
coherence in the sample.

That this is really the case we can see from the comparison of the first and
second column in Table 1. We see that the mean square fluctuation $%
\left\langle \left( \Delta P\right) ^2\right\rangle $ when calculated with
the dispersion relation pertinent for normal electrons namely with ${\cal E}%
_{\alpha m}=[(\hbar \alpha /R-eA_e)^2+\hbar ^2k_z^2(m)]/(2m_e)$ is larger
than $\left\langle \left( \Delta P\right) ^2\right\rangle $ calculated with
the dispertion relation (\ref{pI}), modified by the presence of
self-consistent flux $\phi _I$. Thus the presence of the magnetostatic
coupling decreases fluctuations and the number of normal electrons in the
sample. This is also seen in Fig. 2 where the temperature smearing of the
distribution function calculated with the self-consistent flux is smaller
than that calculated with $\phi _e$. We also checked that, as should be
expected, the mean square fluctuation $\left\langle \left( \Delta P\right)
^2\right\rangle $ decreases with increasing the number of interacting
channels.

This way of analysis bears some resemblance to the two-fluid description of
a superconductor \cite{Enz}. For free electrons with the dispersion relation 
${\cal E}_\alpha $ the electron density can be written as 
\begin{equation}
n=-\frac{2\hbar ^2}{3m_eV}\sum_{\underline{k}}\underline{k}^2\frac{\partial
f({\cal E}\underline{_k})}{\partial {\cal E}_{\underline{k}}}. 
\end{equation}
Defining similarly the density of normal electrons in a superconductor, with
a dispersion relation $E_k=\sqrt{{\cal E}_k+\Delta ^2}$, as 
\begin{equation}
n_n=-\frac{2\hbar ^2}{3m_eV}\sum_{\underline{k}}\underline{k}^2f^{\prime }(E%
\underline{_k}) 
\end{equation}
and the density of super-electrons as 
\begin{equation}
n_c=n-n_n, 
\end{equation}
we get the two-fluid description.

We are going to discuss now the influence of fluctuations on self-sustaining
currents. In Fig. 3 we present the currents given by Eqs. (\ref{pII}), (\ref
{pIII}) at $\phi _e=0$, for different shapes of the FS and for different
values of $\rho $. The fluctuations decrease the current, but
self-consistent solutions can still be obtained for Fermi surfaces with flat
regions ($u=6$ and $u=12$). However, as should be expected, the value of
self-consistent current (flux) is smaller for the case with fluctuations
included. For $u=2$ we do not get flux trapping because the number of
coherent electrons is too small.

\ 

Finally, to get more insight into the properties of our system we discuss,
using the microscopic Hamiltonian (\ref{m11}), the possibility of long-range
order in a cylinder made of a set of mesoscopic rings.

In the literature one finds the statements that phase transitions and
long-range order are impossible in quasi-1D systems. It is true for systems
in the thermodynamic limit with short-range interactions. However, one can
look for the conditions for an ordered state for large but finite number of
interacting entities \cite{Imry}.

Let us consider a set of $M_z$ rings described by the Ising-like Hamiltonian
(\ref{m11}). In the ground state all currents run parallel. Let us construct
a new configuration by reversing the direction of $L$ currents ($L\ll M_z$)
in $s$ ($1\leq s\ll M_z$) different places in the chain of rings. The energy
change will be denoted by $s\Delta E_L(M_z)$, and the change in the free
energy is 
\begin{equation}
\Delta F=s\Delta E_L(M_z)-T\Delta S=s\left[ \Delta E_L(M_z)-k_BT\ln
M_z\right] . 
\end{equation}
If $\Delta F>0$ then the ordered configuration is stable. This condition is
equivalent to 
\begin{equation}
\label{+}\xi _T\equiv e^{\Delta E_L/k_BT}>M_z, 
\end{equation}
where $\xi _T$ has the sens of a (dimensionless) correlation range.

If Eq. (\ref{+}) is fulfilled the system is ordered in the sense that the
correlations extend over all its length.

Let us discuss the possibility of long-range order for the case considered
by us. The energy $\Delta E_L(M_z)$ is of the form: 
\begin{equation}
\Delta E_L(M_z)=\frac 12\left( \frac{e\hbar N^R}{2\pi m_eR^2}\right)
^2\sum_{n/1}^L\ \sum_{m-m^{\prime }/L+1}^{M_z-n}{\cal L}(m-m^{\prime }),
\end{equation}
The $L$ dependence of $\Delta E_L(M_z)$ is presented in Fig. 4. We see that $%
\Delta E_L(M_z)$ increases with $L$ and decreases with $b\equiv
z_{m,m+1}=z_{m+1}-z_m$. The smallest energy change is obtained when
reversing a direction of a single current in several different places.
In Fig. 5 we present the $M_z$ dependence of $\Delta E_1$. We see that
$\Delta E_1$ increases with $M_z$ for small $M_z$ and then saturates.

We are in position now to calculate the temperature $T^{*}$ at which the
crossover from an ordered to disordered state occurs: 
\begin{equation}
\label{pIV}T^{*}=\frac{\Delta E_1(M_z)}{k_B \ln M_z}.
\end{equation}
The calculations performed for the following set of parameters: $M_z=10^4$, $%
M_r=1$, $R=5000$\AA , $b=10$\AA\ gave us $T^{*}\sim $ $0.14K$. It means that
at $T<T^{*}$ the system exhibits a long-range order in the sense that the
correlation range is longer than the sample size. The temperature $T_c$
calculated in the MFA for the above set of parameters is $T_c\sim 0.216K$. 
\cite{Wohll}

We see that $T^{*}$ obtained by the use of the Ising model is of the same
order as $T_c$ obtained with the MFA. It means that long-range interactions
encountered in our system strongly supress fluctuations. If we assume that
each ring has a small number of transverse channels $M_r$ then e.g. for $%
M_r\sim 3$, $T^{*}\sim 1K$.

It has to be stress e.g. that finite value of $T^{*}$ for our system is a
finite size effect. Indeed, as we can see from Eq. (\ref{pIV}) and Fig. 5 $%
T^{*}\rightarrow 0$ for $M_z\rightarrow \infty $.

\section{Discussion and conclusions}

In the presented paper we discussed the magnetostatic coupling of electrons
in mesoscopic systems and its approximations. This interaction is known to
be weak in macroscopic samples, however it seems it can play an important
role in mesoscopic samples due to very peculiar properties of mesoscopic
systems in the magnetic field.

It is possible to induce persistent currents (or in other words orbital
magnetic moments) in mesoscopic ring by the static magnetic field. The
magnetic interaction of orbital magnetic moments can lead to magnetically
ordered ground state. This possibility has been discussed in a number of
papers \cite{Wohll,Fried} using mean field approximation which
states that each electron moves in an external magnetic field and the field
coming from all currents in a system. The obtained two self-consistent
equations for the current can lead to spontaneous self-sustaining current at
zero external field. We neglected here the Zeeman energy of the electron
spins because it turns out to be very small compared to the orbital
energies. The influence of spins has been discussed in Ref. \cite{ZipSzop}.

In this paper we have presented the microscopic Hamiltonian which is
responsible for the internal magnetic field - the magnetostatic
(current-current) interaction. The strength of the interaction depends
strongly on the sample geometry. For the stack of mesoscopic rings
deposited along certain axis we get the long-range interaction with the
coupling constant depending only on the radii of the rings and on the
relative distance of its centers.

We have shown that the self-consistent MFA of the current-current
interaction, gives the effective Hamiltonian $H^{MF}$ leading to
self-sustaining currents. Its derivation from the long-range interaction
serves as a justification of the use of the Hamiltonian (\ref{m28}) to
investigate magnetic properties of mesoscopic systems. The MFA is known to
be the best for systems with long-range forces (if the condition (\ref{mS})
is fulfilled), thus we should expect that it leads to reasonable results in
the considered case.

To obtain the full Hamiltonian describing our system one should add to
Hamiltonian given by Eq. (\ref{m19}) the Coulomb interaction and the
interaction with impurity potential. It was recently shown that the Coulomb
interaction does not influence persistent currents in clean systems \cite
{AviBrav}, whereas it enhances the current in diffusive regime \cite
{BerkAvi}. In a work by M. Pascaud and G. Montambaux \cite{Pasc} the
experiments which permit to test the role of Coulomb interaction have been
suggested.

In the model calculations presented in this paper we did not consider the
effect of impurities in order not to obscure the whole subject with too
many details. The influence of disorder on self-sustaining currents has
been analysed in Ref. \cite{Lis}. We found that disorder decreases
persistent currents but self-sustaining currents can still be obtained
for relatively clean samples (ballistic regime).

To go beyond the MFA we have considered the influence of fluctuations,
calculated by the use of Eq. (\ref{b24}), on the properties of a mesoscopic
cylinder. We have shown that these fluctuations are smaller in mesoscopic
systems than in macroscopic ones because of the quantum size energy gaps. On
the top of it the magnetostatic coupling modifies a dispertion relation and
creates a dynamic gap what leads to further reduction of the fluctuations.
Thus in mesoscopic systems coherent and normal electrons coexist and the
system can be described by the two-fluid model where the fluctuations are
proportional to the amount of normal electrons. Self-sustaining currents run
by coherent electrons survive fluctuations in systems with FS having flat
regions, however their magnitude is reduced.

Having the microscopic Hamiltonian for electrons interacting by
magnetostatic coupling (Eq. (\ref{m11})) we have discussed, for a set of
stacked rings, the possibility of long-range order. Phase transitions and
long-range order are possible in the strict mathematical sense only for
systems in the thermodynamic limit. For finite (but still large) systems
discontinuities of thermodynamic quantities and infinite range correlations
are not necessary. ''Discontinuities'' have finite widths and correlation
ranges may be as large as the system itself, regardless of the behaviour in
the thermodynamic limit \cite{Imry}. What is more finite systems can show
interesting effects which will be wiped away in the thermodynamic limit.
This is the situation in the presented paper where at temperatures $T\leq
T^{*}\sim 0.1K-1K$ the set of mesoscopic rings (or in general the mesoscopic
cylinder with quasi-1D conduction) can exhibit the long-range order, but $%
T^{*}\rightarrow 0$ for $M_z\rightarrow \infty $.

The statistical properties of the system following from Eq. (\ref{m11}) will
be presented in a subsequent paper.

\section{Acknowledgements}

We would like to thank W. Brenig, Y. Imry, W. Zwerger for usefull
discussions. Work was supported by KBN Grant 2P03B 129 14 and by DAAD Grant.

\begin{figure}
\caption{The interaction constant ${\cal L}$ as a function of
distance $z$ between the ring centres.}
\end{figure}

\begin{figure}
\caption{The Fermi-Dirac distribution function $f(E_\alpha )$ versus
energy $E_\alpha $, for the cylinder made of a set of quasi-1D mesoscopic
rings, in the magnetic flux $\phi _e$ and $\phi =\phi _e+\phi _I$
respectively. $\phi_I$ has been calculated in the self-consistent way for
parameters as in the figure.}
\end{figure}

\begin{figure}
\caption{Persistent currents $I$ as a function of magnetic flux $\phi /\phi
_0 $ in quasi-2D mesoscopic cylinders with different shapes of the Fermi
surfaces, with and without fluctuations. Self-sustaining currents $I_s$.}
\end{figure}

\begin{figure}
\caption{The energy change $\Delta E_L/W$, where $W=\frac 12\left( \frac{e
\hbar N^R}{2\pi m_eR^2}\right) ^2$, as a function of the number of
magnetic moments $L$ for different values of $b\equiv z_{m,m+1}$.}
\end{figure}

\begin{figure}
\caption{The energy change $\Delta E_1$ and the temperature $T^{*}$ as
a function of the number of channels $M_z$ in a single cylinder.}
\end{figure}

\newpage

\begin{table}
\centering
  \begin{tabular}{|l|l|l|l|l|l|l|} \hline
{\bf Shape of the FS} & {$\phi_e /\phi_0$} & {$\phi/\phi_0$} &
{$\rho^{\phi_e}$} & {$\rho^{\phi}$} &
{${N_n}^{\phi_e}$ x$10^6$} & {${N_n}^{\phi}$ x$10^6$} \\ \hline
half-circular     &   0.000   &  0.000  &  0.9851  &  0.9851  &
  9852.22  &  9852.22  \\
(u=2)             &   0.100   &  0.088  &  0.9968  &  0.9944  &
  9969.15  &  9944.78  \\ \hline
rectangular with  &   0.000   &  0.000  &  0.0695  &  0.0695  &
  0695.52  &  0695.52  \\
rounded corners   &   0.100   &  0.017  &  0.1207  &  0.0709  &
  1207.51  &  0708.92  \\
(u=6)             &           &         &          &          &
           &           \\ \hline
rectangular       &   0.000   &  0.000  &  0.0688  &  0.0688  &
  0687.00  &  0687.00  \\
(u=12)            &   0.100   &  0.016  &  0.1195  &  0.0700  &
  1195.00  &  0700.00  \\
                  &   0.250   &  0.041  &  0.5725  &  0.0767  &
  5727.00  &  0768.00  \\ \hline
  \end{tabular}
  \
  \caption{The relative mean square fluctuations of the total momentum $P$,
$\rho={<(\Delta P)^2>}/N m_e k_B T$,
in a system made of 2D coaxial cylinders with $N$=10001x$10^6$ conducting
electrons (lattice constant $a$=1\AA, $b$=5\AA) at temperature $T$=15K,
in the magnetic flux $\phi_e$ and $\phi=\phi_e + \phi_I$
respectively, for different shapes of the Fermi surface. $\phi_I$ has been
calculated for each $\phi_e$ in the self-consistent way, assuming the number
of interacting coaxial cylinders $M_r=100$. $N_n$ represent the total number
of ''normal'' electrons in the system. The number of coherent electrons is
$N_c = N - N_n$.}
\end{table}

\end{document}